\documentclass{article}
\usepackage[margin=1in]{geometry}
\usepackage{amsmath, amssymb, graphics, setspace}
\usepackage{tikz}
\usepackage[style=ieee, citestyle=numeric-comp, backend=biber, url=false]{biblatex} 
\usepackage{multirow}
\usepackage{caption}
\usepackage{subcaption}
\usepackage{graphicx}

\newcommand{\mathsym}[1]{{}}
\newcommand{\unicode}[1]{{}}
\usepackage{boldline}

\title{Reheating constraints on mutated hilltop inflation}
\author{Sudhava Yadav$^1$, Dhwani Gangal$^1$ and K.K. Venkataratnam$^{1,\footnotemark{}}$  
\\$^1$ Department of Physics, Malaviya National Institute of Technology Jaipur, Jaipur-302017, India}
\begin{document}
\maketitle
\begin{abstract}
   Future research studies of cosmic microwave background polarization seems likely to provide a more improved upper bound of $r \le 0.03$ on the tensor-to-scalar ratio(r). In our work, we have done the reheating study of mutated hilltop inflation(MHI), a model falling in the broad category of small field inflation. We have parameterized reheating in terms of various parameters like reheating duration $N_{\text{rh}}$, reheating temperature $T_{\text{rh}}$ and effective equation of state \(\overline{\omega }_{\text{rh}}\) using observationally viable values of  scalar power spectrum amplitude \(A_{\text{s}}\) and scalar spectral index \(n_{\text{s}}\). In our study, working over a range of \(\overline{\omega }_{\text{rh}}\), we found that the MHI potential is well consistent with combined Planck and BK18 observations for  $\overline{\omega }_{\text{rh}} > 0$ within a particular range of model's parameter space and the lower values of the model parameter in MHI generate considerably smaller r compared to normal hilltop potential without any incompatibility of $n_s$ with observational data, making MHI a better choice in accordance to recent and future studies. 
\end{abstract}
\footnotetext{*Corresponding author}
\begin{section}{Introduction}\label{S1}
Inflation is a well recognized theory for explaining the early universe physics\cite{guth_inflationary_1981,starobinsky_new_1980,linde_new_1982,linde_chaotic_1983,riotto_inflation_2017,riotto_inflation_2002}, and it has addressed multiple cosmological issues. By assuming a brief era of accelerated expansion at an early stage of universe, it provides a logical explanation for genesis of observed structures, homogeneity, and flatness of universe.
In its most basic form, inflation is ruled by a scalar field called the inflaton $\phi$, evolving slowly through an almost flat potential $V(\phi)$. The quantum fluctuations in inflaton field give rise to perturbations in primitive stage of universe and are seen in cosmic microwave background (CMB) radiation as temperature fluctuations \cite{mukhanov_quantum_1981,starobinsky_dynamics_1982,guth_quantum_1985}. Inflation predicts almost scale-invariant gaussian spectrum of early universe density perturbations.\\
In general, the equations governing motion of inflaton are analytically unsolvable without making certain approximations. The slow roll approximation\cite{Albrecht:1982wi, linde_new_1982}, is robust one that considers inflaton’s kinetic energy being dominated by the expansion making it possible to find analytic solutions to the equations governing motion of inflaton and to express the primordial power spectrum using the slow roll parameters.\\
While the inflationary phase has got substantial support, the details of end of this phase and transition to later stages of universe is still an active research area. Reheating\cite{albrecht_reheating_1982,kofman_reheating_1994,kofman_towards_1997,drewes_kinematics_2013,allahverdi_reheating_2010}, an epoch serving as a link between inflationary and radiation dominated eras, initiates the thermalization process and governs the ensuing evolution of observable universe. The inflation ends with a state of universe that is highly non-thermal and is thermalized later by scattering, creating a blackbody spectrum in the universe at a temperature (\(T_{\text{rh}}\)). This is the temperature at end of reheating or onset of radiation dominance. Another important parameter is the duration of reheating, defined by number of e-folds (\(N_{\text{rh}}\)) from end of inflation to the onset of radiation dominance. During reheating, the energy density evolution of cosmic fluid depends on a parameter known as effective equation of state(EoS) (\(\overline{\omega }_{\text{rh}}\)), which takes the values (-1/3 to 1 ) during different epochs.\\
There is a lack of well-established science supporting reheating, but the recent CMB observations made it possible to determine indirect constraints on various reheating parameters\cite{martin_observing_2015,martin_first_2010,dai_reheating_2014,martin_inflation_2006,adshead_inflation_2011,mielczarek_reheating_2011,cook_reheating_2015,goswami_reconciling_2018,yadav_reheating_2023}. The equation relating the reheating parameters \(T_{\text{rh}}\), \(N_{\text{rh}}\) and \(\overline{\omega }_{\text{rh}}\) with spectral index \(n_{\text{s}}\) can be derived and used to bound \(n_{\text{s}}\), r and \(N_{\text{k}}\) by demanding \(\overline{\omega }_{\text{rh}}\) varying in the range ($-\frac{1}{3} \le \overline{\omega }_{\text{rh}} \le 1$) along with the condition \(T_{\text{rh}} >\) 100 GeV for dark matter production at weaker scales.\\
Now, talking about the conditions of flatness, one way out is inflation happening near maxima of potential. This situation is called ‘Hilltop’ inflation. The advantage of this scenario is that it is easy to satisfy the slow roll conditions. The hilltop inflationary potentials\cite{boubekeur_hilltop_2005,tzirakis_inflation_2007} are a part of widely studied single- field inflationary models \cite{ben-dayan_cosmic_2010,Hotchkiss_2012,Antusch_2014,Garcia_Bellido_2014,wolfson_small_2018,wolfson_likelihood_2019}. Many variants of hilltop inflation are presented in literature\cite{dimopoulos_analytic_2020,lin_type_2019,kallosh_hilltop_2019,kohri_more_2007,lin_super_2009}; one of them is mutated hilltop inflation \cite{pal_mutated_2010} where a hyperbolic function having power series expansion containing an infinite number of terms is added to the flat potential.
In this work, we will be doing the reheating study of both normal and mutated hilltop models. We used Planck 2018 bound on $n_s$\cite{aghanim_planck_2020,akrami_planck_2020-1} and combined Planck and BK18 bound on r, i.e. (\(r<0.032)\)\cite{tristram2022improved} to put reheating constraints on these models. Furthermore, subsequent measurements by BICEP\cite{bicep2_collaboration_detection_2014,ade_bicep2_2014,keck_array_and_bicep2_collaborations_constraints_2018,Barkats_2014,wu_initial_2016} might lower this upper bound on r from $r \le 10^{-2}$ to $r \le 10^{-3}$ \cite{easther_running_2022,abazajian_cmb-s4_2016}.
\\
The organization of this article is as follows: In Sec. 2, we have briefly reviewed the reheating formalism and presented the equations for \(N_{\text{rh}}\)
and \(T_{\text{rh}}\) in terms of \(\overline{\omega }_{\text{rh}}\), \(\text{N}_k\) and \(V_{\text{end}}\). In Sec.3, we have done the reheating study for normal hilltop and mutated hilltop inflation using the variation of \(T_{\text{rh}}\) and \(N_{\text{rh}}\) for these models with \(n_s\) for varied choices of \(\omega _{\text{rh}}\) to put reheating constraints on the models using BK18 and Planck 2018 data\cite{aghanim_planck_2020,akrami_planck_2020-1,tristram2022improved,ade2021improved}. Sec.4 contains our discussion and conclusion.
\end{section}
\begin{section}{Reheating Formalism}\label{S2}
To establish our notation, we quickly go over the reheating formalism proposed in ref. \cite{goswami_reconciling_2018,yadav_reheating_2023}.
For single-field models of inflation, the inflaton($\phi$) with potential $V(\phi)$ emerges slowly with parameters given as
\begin{equation} \label{4}
H^2=\frac{V}{3M_P^2},
\end{equation}
% where \((')\) denotes differentiation w.r.t \(\phi\). 
% The slow-roll parameters during inflation are defined as
\begin{equation} \label{5o}
\epsilon =\frac{1}{2}\left(M_P\frac{V'}{V}\right)^2,
\end{equation}
\begin{equation} \label{5}
\eta =M_P^2\left(\frac{V''}{V}\right).
\end{equation}
 where H is the Hubble parameter and \((')\) denotes the \(\phi\) derivative. Now, using these parameters, tensor spectral index \(n_T\), scalar spectral index \(n_s\) and tensor to scalar ratio \(r\) can be expressed as

\begin{equation}\label{6}
n_T=-2\epsilon.
\end{equation}

\begin{equation}\label{7}
\text{   }n_s=1-6\epsilon +2\eta.
\end{equation}

\begin{equation}\label{8}
r=16\epsilon.
\end{equation}
The e-folds between the termination of inflation and the moment at which a mode k passes the Hubble horizon, \(\text{N}_k\), are
\begin{equation}\label{9}
N_k=\frac{1}{M_P^2}\int_{\phi _{\text{end}}}^{\phi _k} \frac{V}{V'} \, d\phi,
\end{equation}
Where \(\phi _{k }\) is the value $\phi$ takes when k undergoes Hubble crossing. Let us assume an energy density controlling the evolution of universe during reheating and characterized by 
\begin{equation} \label{10}
\text{    }\rho _{\text{rh}}=\frac{\pi ^2}{30}g_{\text{rh}}T_{\text{rh}}^4,
\end{equation}
where \(g_{\text{rh}}\) is the relativistic species count at reheating end and \(T_{\text{rh}}\) is the reheating temperature. We have taken \(g_{\text{rh}}\) $\approx $ 100 \cite{dai_reheating_2014}. Now, defining \(\overline{\omega }_{\text{rh}}\) as the  effective EoS parameter during reheating and \(a_{rh}\) as the scale factor at the end of reheating, we can express number of e-foldings during reheating as
\begin{equation} \label{11}
N_{\text{rh}}= \frac{a_{\text{rh}}}{a_{\text{end}}}= -\frac{1}{3(1+\overline{\omega
}_{\text{rh}})}\ln\left(\frac{\rho _{\text{rh}}}{\rho _{\text{end}}}\right).
\end{equation}
Rewriting \(N_{\text{rh}}\) using eq. (\ref{10})
\begin{equation} \label{12}
N_{\text{rh}}=\frac{1}{3\left(1+\overline{\omega }_{\text{rh}}\right)}\left\{\ln \left(\frac{3}{2}V_{\text{end}}\right)-\ln \left(\frac{\pi ^2}{30}g_{\text{rh}}\right)\right\}-\frac{4}{3\left(1+\overline{\omega
}_{\text{rh}}\right)}\ln \left(T_{\text{rh}}\right).
\end{equation}
 Now, the wavenumber `\(\frac{k}{a}\)' for a physical scale \(k\), can be given in terms of above introduced quantities as
\begin{equation} \label{13}
H_k=\frac{k}{a_k}=\left(1+z_{\text{eq}}\right)\frac{k}{a_o}\rho _{\text{rh}}^{\frac{3~\overline{\omega }_{\text{rh}}-1}{12\left(1+\overline{\omega
}_{\text{rh}}\right)}}\rho _{\text{eq}}{}^{-\frac{1}{4}}\left(\frac{3}{2}V_{\text{end}}\right)^{\frac{1}{3\left(1+\overline{\omega }_{\text{rh}}\right)}}e^{N_k},
\end{equation}
where (\(z_{\text{eq}}\)) is the redshift during the epoch of matter-radiation equality and we have taken \(z_{\text{eq}}=\) 3402 \cite{aghanim_planck_2020,,akrami_planck_2020-1}. Solving eq. (\ref{13}) for \(N_k\) gives
\begin{equation} \label{14}
N_k=\ln  H_k-\ln \left(1+z_{\text{eq}}\right)-\ln \left( \frac{k}{a_o}\right)-\frac{1}{\left.3(\overline{\omega
}_{\text{rh}}+1\right)}\ln \left(\frac{3}{2}V_{\text{end}}\right)-\frac{3~\overline{\omega }_{\text{rh}}-1}{3\left(1+\overline{\omega
}_{\text{rh}}\right)}\ln \left(\rho _{\text{rh}}^{\frac{1}{4}}\right)+\ln \left(\rho _{\text{eq}}^{\frac{1}{4}}\right).
\end{equation}
By making use of eq. (\ref{14}) and  eq. (\ref{10}), a mutual relation is obtained among different parameters introduced,
\begin{equation} \label{15}
\ln  (T_{\text{rh}})=\frac{3~\left(1+\overline{\omega }_{\text{rh}}\right)}{3~ \overline{\omega }_{\text{rh}}-1}\left\{\ln  H_k-\ln
\left(1+z_{\text{eq}}\right)-\ln  \frac{k}{a_o}- N_k+\ln \left(\rho _{\text{eq}}^{\frac{1}{4}}\right)\right\}
-\frac{1}{3~\overline{\omega }_{rh}-1}\ln \left(\frac{3}{2}V_{end}\right)-\frac{1}{4}\ln \left(\frac{\pi ^2}{30}g_{rh}\right).
\end{equation}
The expression of \(T_{\text{rh}}\) obtained by rearranging eq. (\ref{12}) is inserted in eq. (\ref{15}) to get an expression for \(N_{\text{rh}}\) as given below
\begin{equation} \label{16}
N_{\text{rh}}=\frac{1}{3~\overline{\omega }_{\text{rh}}-1}\ln \left(\frac{3}{2}V_{\text{end}}\right)+\frac{4}{3~\overline{\omega }_{\text{rh}}-1}\left\{\ln
\left(\frac{k}{a_o}\right)+ N_k+\ln \left(1+z_{\text{eq}}\right)-\ln \left(\rho _{\text{eq}}^{\frac{1}{4}}\right) -\ln  H_k \right\}.
\end{equation}
eq. (\ref{15}) and eq. (\ref{16}) are two important relations which parameterise 
 the reheating epoch.
\end{section}
\begin{section}{Small field inflationary models(SFI)}\label{S3}
    \begin{subsection}{Hilltop Inflation(HI)}\label{S3.1}
Hilltop inflation occurs nearby maxima of a sufficiently flat potential having the form \cite{boubekeur_hilltop_2005,tzirakis_inflation_2007}
\begin{equation}\label{17}
V = M^4[1- \left(\frac{\phi}{\mu}\right)^p].
\end{equation}
where M is the normalization term, $\mu$ is the mass scale and p is the power index. This was considered in the context of reheating in \cite{cook_reheating_2015}. We are just quickly reproducing the results using recent Planck+BK18 observational data before moving to mutated hilltop inflation. We are considering $p \ge 2$ for our analysis.
\begin{figure}[!h]
 \centering
\includegraphics[width=0.48\textwidth]{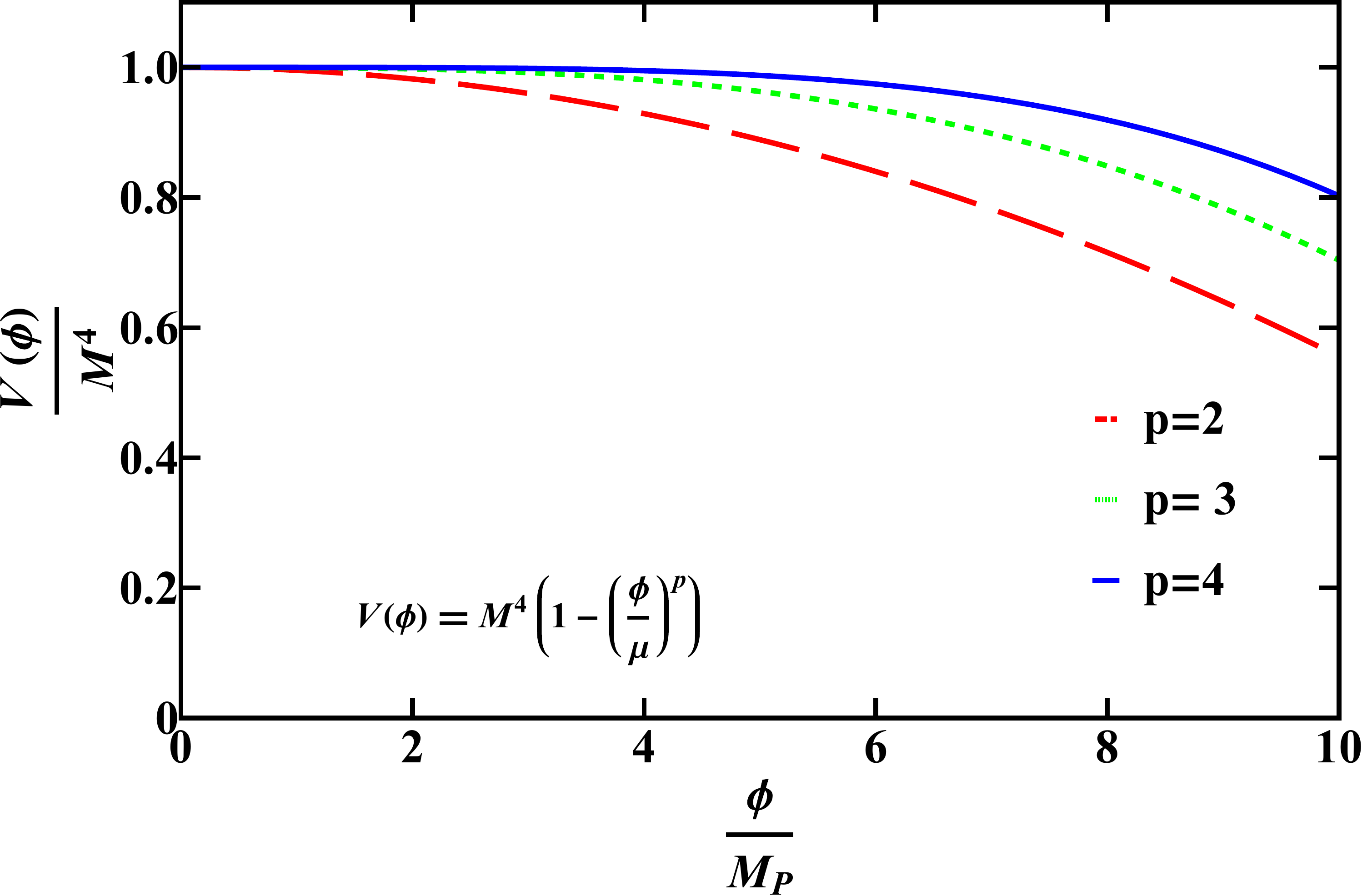}
    \caption{Potential versus $\frac{\phi}{M_P}$ for normal hilltop inflation \(V = M^4[1- \left(\frac{\phi}{\mu}\right)^p]\) for ${\mu}=15M_P$ and different values
of {p} : {p}=2 (large dashed red), {p}=3 (tiny dashed green), {p}=4 (solid blue).} 
\label{F1}
\end{figure}\\
Using eqs. (\ref{4} ) and (\ref{5} ), the hubble parameter and slow-roll parameters for hilltop potential can be expressed as\\
\begin{equation}\label{18}
H^2 = \frac{M^4[1- \left(\frac{\phi}{\mu}\right)^p]}{3M_P^2},
\end{equation}  

 \begin{equation}\label{19}
\text{$\epsilon $ =}\frac{M_P^2p^2 \left(\frac{\phi }{\mu}\right)^{2 p} }{2 \phi ^2 \left(-1+\left(\frac{\phi }{\mu}\right)^p\right)^2},
\end{equation}

\begin{equation}\label{20}
\text{$\eta $ = }\frac{M_P^2(-1+p) p \left(\frac{\phi }{\mu}\right)^p }{\phi ^2 \left(-1+\left(\frac{\phi }{\mu}\right)^p\right)},
\end{equation}
where \(M_{P}\) is the reduced Planck{'}s mass having value \(M_{P}=\text{}\)2.435 $\times
$ \(10^{18}\) GeV. Now, considering a mode \(k_*\) same as Planck collaboration, \(\frac{k_*}{a_o}=0.05 Mpc^{-1}\), whose hubble crossing happens at field value \(\phi _*\). The number of e-folds left after pivot scale \(k_*\) crosses the Hubble radius are
\begin{align}\label{21}
N_* & \simeq \frac{1}{M_P^2}\int_{\phi _{\text{end}}}^{\phi _*} \frac{V}{V'} \, d\phi, \nonumber\\& = \frac{\frac{\phi _*^2-\phi _{\text{end}}^2}{2}-\frac{\phi _*^2 \left(\frac{\phi _*}{\mu}\right)^{-p}-\phi _{\text{end}}^2 \left(\frac{\phi _{\text{end}}}{\mu}\right)^{-p}}{2-p}}{p
M^2}\qquad(for\quad p\not= 2)\nonumber\\& = \frac{\phi _*^2-\phi _{\text{end}}^2+2\mu^2 \left(ln(\frac{\phi _{\text{end}}}{\mu})-ln(\frac{\phi _*}{\mu})\right)}{4
M^2}\qquad(for\quad p = 2).
\end{align}

  \begin{figure}
      \centering
\includegraphics[width=\textwidth]{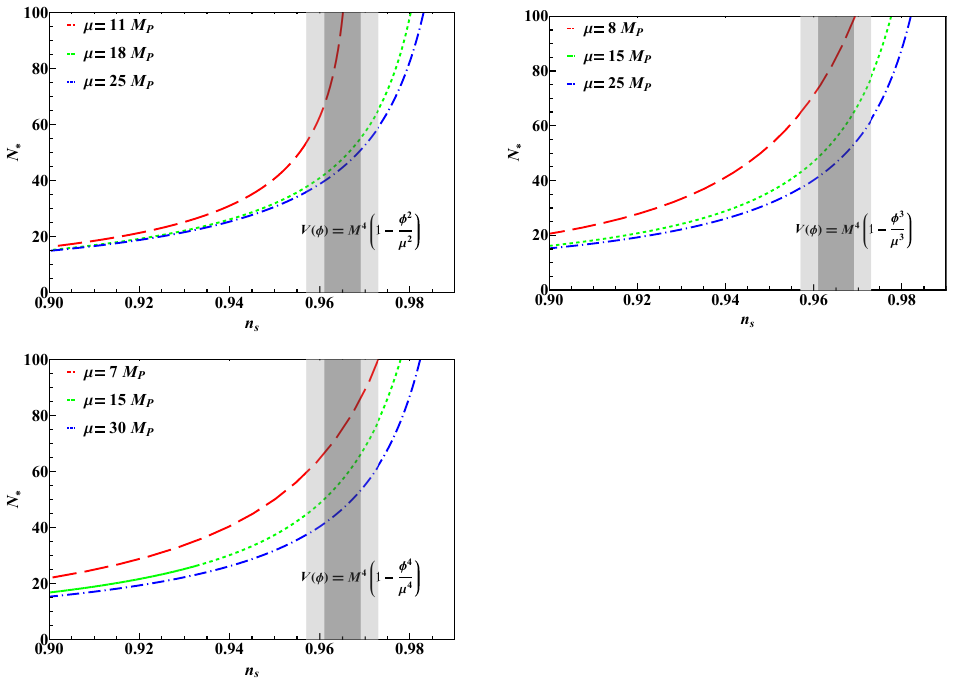}
      \caption{The plots showing variation of \(N_*\) w.r.t. \(n_{s }\) for normal hilltop inflation $V = M^4[1- \left(\frac{\phi}{\mu}\right)^p]$ for various choices of ${\mu}$. The light and dark shades of grey colour depicts the \(\text{2$\sigma $}\) and \(\text{1$\sigma $}\) bounds on \(n_s\) from Planck 2018 (TT+TE+EE+Low E+Lensing) data \cite{aghanim_planck_2020,,akrami_planck_2020-1}.}
      \label{F2}
  \end{figure}
Using the eqs. (\ref{19}) and (\ref{20})\ for $\epsilon $ and $\eta $ in eq. (\ref{7}) and  eq. (\ref{8}), the scalar spectral index and tensor-to-scalar ratio \(r=16\epsilon\) can be given as
\begin{equation}\label{22}
n_s=1-\frac{M_P^2p \left(\frac{\phi _*}{\mu}\right)^p \left(2 \left(\left(\frac{\phi _*}{\mu}\right)^{p} -1 \right)+p \left(2+\left(\frac{\phi _*}{\mu}\right)^p\right)\right)
}{\phi _*^2 \left(\left(\frac{\phi _*}{\mu}\right)^{p} -1\right)^2}.
\end{equation}
\begin{equation}\label{23}
r=\frac{8M_P^2p^2 \left(\frac{\phi _*}{\mu}\right){}^{2 p} }{ \phi _*^2 \left(1-\left(\frac{\phi _*}{\mu}\right)^{p}\right)^2}.
\end{equation}
The variation of \(N_*\) with \(n_{s\text{  }}\) for p = 2,3 and 4 with three different choices of $\mu$ taken in each case are plotted in figure (\ref{F2}). The light and dark shades of grey colour depicts the \(\text{2$\sigma $}\) and \(\text{1$\sigma $}\) bounds on \(n_s\) from Planck's 2018 data (TT+TE+EE+Low E+Lensing) \cite{aghanim_planck_2020,,akrami_planck_2020-1}.\\
Moreover, this model yields the relation
\begin{equation}\label{24}
H_*=\pi \text{  }M_P^2\sqrt{8A_s\frac{p^2 \left(\frac{\phi _*}{\mu}\right)^{2 p} }{2 \phi _*^2  \left(-1+\left(\frac{\phi _* }{\mu}\right)^p\right)^2}}.
\end{equation}
This, along with eq. (\ref{18}) and the condition governing the inflationary end, gives the value \(V_{\text{end}}\),
\begin{equation}\label{25}
V_{\text{end}}= 3H_*^2 M_P^2 \frac{1- \left(\frac{\phi _{\text{end}}}{\mu}\right)^p}{1-\left(\frac{\phi _*}{\mu}\right)^p}.
\end{equation}
Now, \(N_*\), $n_s$, \(H_{\text{*}}\) and \(V_{\text{end}}\) from eqs. (\ref{21}), (\ref{22}), (\ref{24}) and (\ref{25}) are all inserted in eqs. (\ref{15}) and (\ref{16}) to get plots of \(N_{\text{rh}}\) and \(T_{\text{rh}}\) as a function of \(n_s\)  for different power indices(p). These plots for p = 2, 3 and 4 with three different values of $\mu$ taken in each case are shown in figures (\ref{F3}), (\ref{F4}) and (\ref{F5}) respectively along with
Planck-2018 \(1\sigma\) bound \(n_s\text{=0.965$\pm $0.004}\) (light grey) and \(2\sigma\) bound \(n_s\text{=0.965$\pm $0.008}\) (dark grey). We have used \(\text{
 }A_{s }=2.139\times 10^{-9}\). The figures (\ref{F3}), (\ref{F4}) and (\ref{F5}) illustrates that irrespective of power index the curves for different \(\overline{\omega }_{\text{rh}}\) shifts outside the observational bounded region and shift towards lower $n_s$ as the $\mu$ value decreases.\\
 Moving further demanding \(T_{\text{rh}} \ge 100\) GeV, we have obtained allowed range of \(n_s\) and reflected it on eq. (\ref{21}) and (\ref{23}) to get bounds on \(N_*\) and r for different power indices and are presented in table \ref{T1}, \ref{T2} and \ref{T3}. The r versus $n_s$ predictions using these tables for 3 different power indices and with different $\mu$ over range of  \(\overline{\omega }_{\text{rh}}\) values are shown in figure \ref{F6}. It can be seen from figure \ref{F6} that there is a range of $\mu$ values for each power index for which the HI model is consistent with observational data and with increasing values of power index(p) this $\mu$ range becomes wider as can be seen for p=2, $\mu=18 M_P$ is outside the compatible range while for p=4 even $\mu=30 M_P$ shows compatibility with data. It can also be seen that below a minimum $\mu$ there is inconsistency with observational $n_s$ value. As the power index(p) increases the minimum $\mu$ shifts towards lower $\mu$ values. 
 
  \begin{figure}
      \centering
\includegraphics[width=\textwidth]{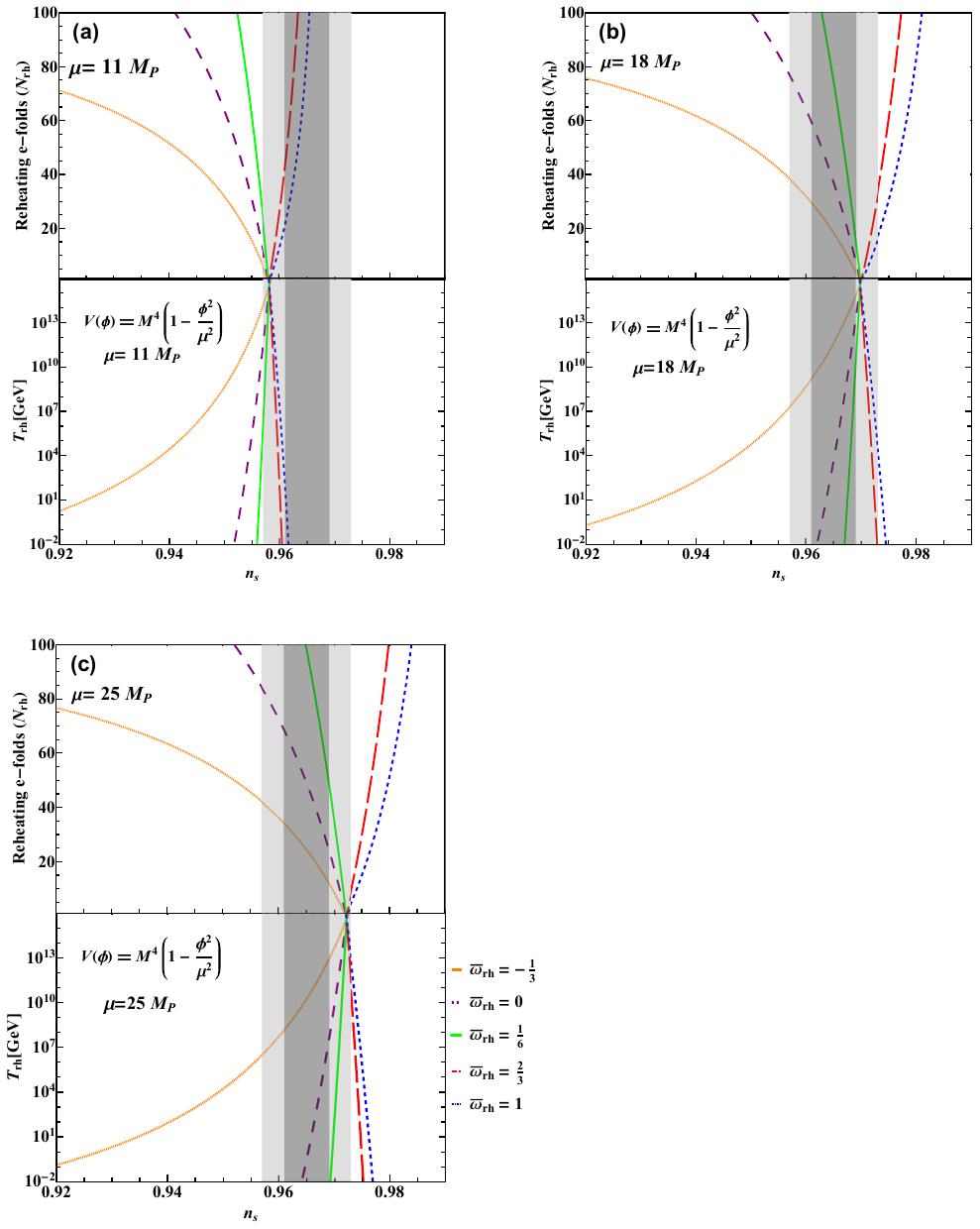}
      \caption{The variation of \(N_{\text{rh}}\) and \(T_{\text{rh}}\)  with spectral index (\(n_{s }\)) for $V = M^4[1- \left(\frac{\phi}{\mu}\right)^2]$ and different choices of ${\mu}$ and \(\overline{\omega }_{\text{rh}}\): \(\overline{\omega }_{\text{rh}}\)=\(-\frac{1}{3}\)(orange dotted), \(\overline{\omega }_{\text{rh}}\)= 0(medium dashed dark purple), \(\overline{\omega
   }_{\text{rh}}\)=\(\frac{1}{6}\)(solid green), \(\overline{\omega }_{\text{rh}}\)=\(\frac{2}{3}\)(big dashed red), \(\overline{\omega }_{\text{rh}}\)= 1(little dashed blue). The grey shading are same as figure \ref{F2}}
      \label{F3}
  \end{figure}
  \begin{figure}
      \centering
 \includegraphics[width=\textwidth]{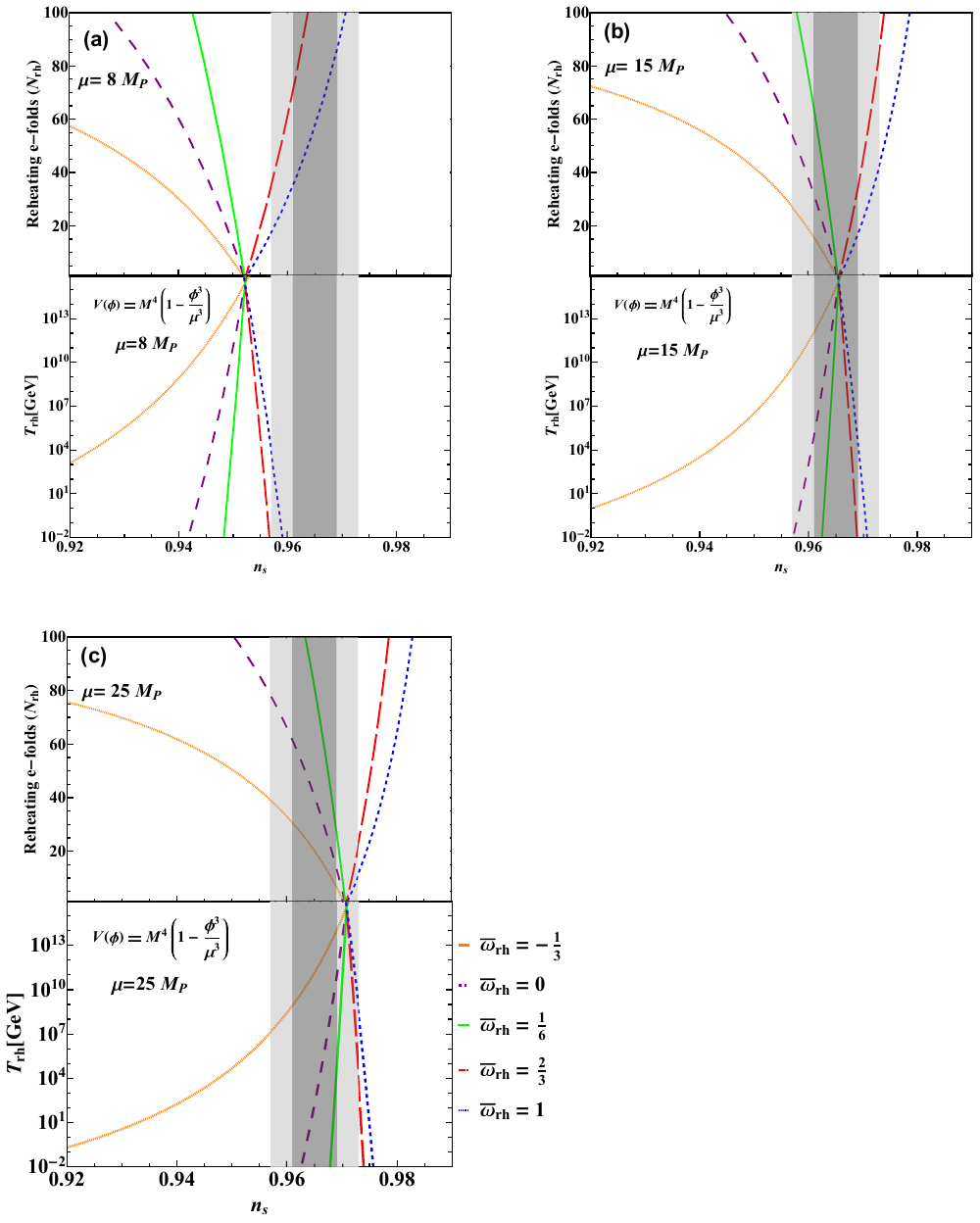}
      \caption{The variation of \(N_{\text{rh}}\) and \(T_{\text{rh}}\)  with spectral index (\(n_{s }\)) for $V = M^4[1- \left(\frac{\phi}{\mu}\right)^3]$ and different choices of ${\mu}$ and \(\overline{\omega }_{\text{rh}}\).The colour codes are identical to figure \ref{F3}}
      \label{F4}
  \end{figure}
  \begin{figure}
      \centering
\includegraphics[width=\textwidth]{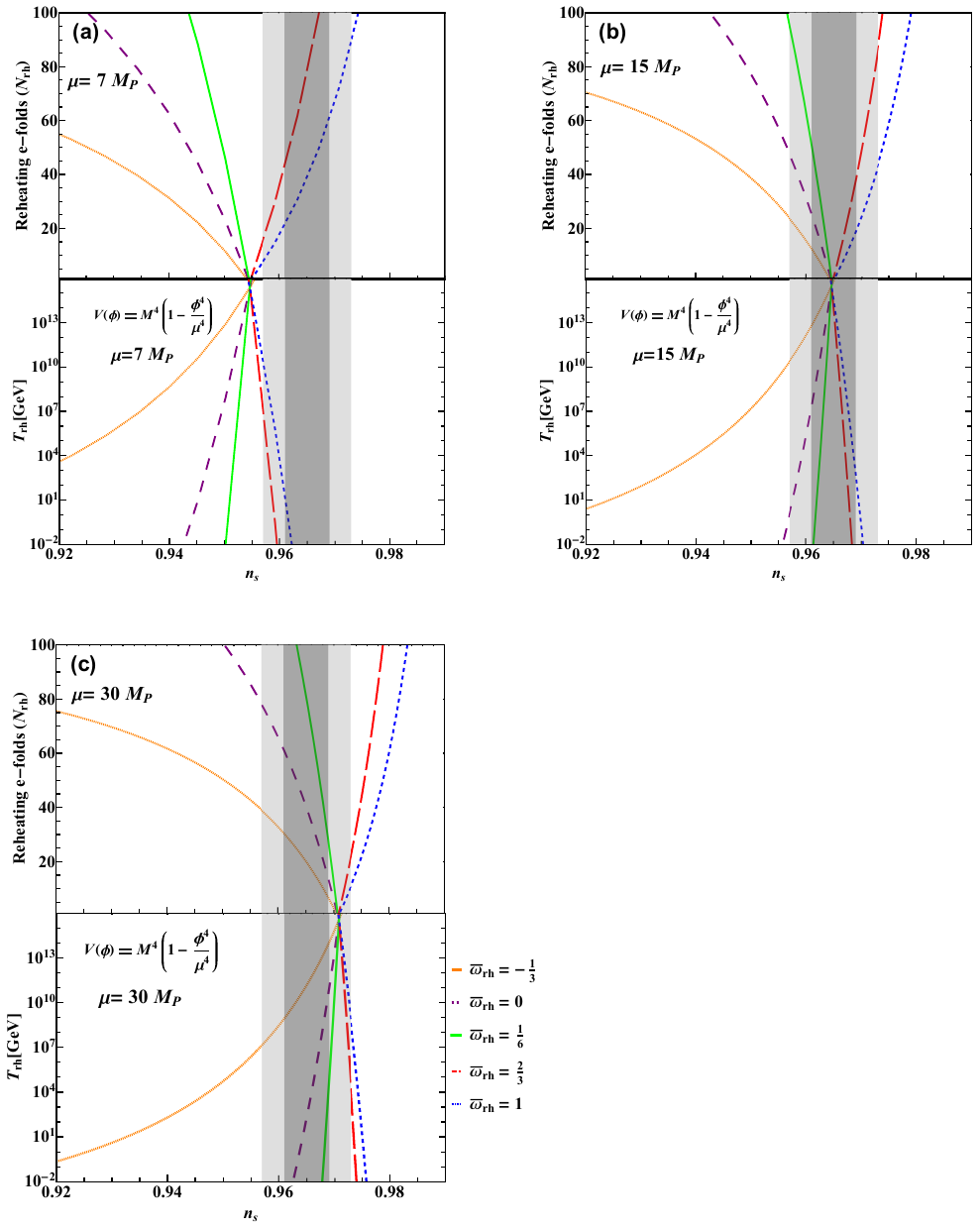}
      \caption{The variation of \(N_{\text{rh}}\) and \(T_{\text{rh}}\)  with spectral index (\(n_{s }\)) for $V = M^4[1- \left(\frac{\phi}{\mu}\right)^4]$ and different choices of ${\mu}$ and \(\overline{\omega }_{\text{rh}}\). The colour codes are identical to figure \ref{F2}}
      \label{F5}
  \end{figure}

\begin{table}
\caption{The allowed ranges of $n_s$, \(N_*\) and $r$ for 3 different choices of ${\mu}$ for normal hilltop inflation by demanding $p=2$ and $T_{rh} \geq 100GeV$}
    \centering
    \begin{tabular}{|c|c|c|c|c|c|}
        \hline
          n = 2 & Effective equation of state & $n_s$ & $ N_*$ & $r$ \\
         \hlineB{4}
         \multirow{5}{*}{u = $11 M_P$}
        & $-\frac{1}{3} \le \overline{\omega }_{rh} \le 0$ & $0.930 \le n_s \le 0.953$ & $25.42 \le N_* \le 46.30$ & $0.077   \ge r \ge 0.027$  \\
         \cline{2-5}
        & $0 \le \overline{\omega }_{rh} \le \frac{1}{6}$ & $0.953 \le n_s \le 0.957$ & $46.30 \le N_* \le 52.21$ & $0.027 \ge r \ge 0.021 $ \\
         \cline{2-5}
        & $\frac{1}{6} \le \overline{\omega }_{rh} \le \frac{2}{3}$ & $0.957 \le n_s \le 0.960$ & $52.21 \le N_* \le 62.78$ & $0.021 \ge r \ge 0.014$ \\
        \cline{2-5}
        & $\frac{2}{3} \le \overline{\omega }_{rh} \le 1$ & $0.960 \le n_s \le 0.961$ & $62.78 \le N_* \le 66.86$ & $0.014 \ge r \ge 0.012$ \\
        \hlineB{4}
        \multirow{4}{*}{u = $18 M_P$}
        & $-\frac{1}{3} \le \overline{\omega }_{rh} \le 0$ & $0.939 \le n_s \le 0.964$ & $25.52 \le N_* \le 46.49$ & $0.110 \ge r \ge 0.050 $ \\
         \cline{2-5}
        & $0 \le \overline{\omega }_{rh} \le \frac{1}{6}$ & $0.964 \le n_s \le 0.968$ & $46.49 \le N_* \le 52.43$ & $0.050  \ge r \ge 0.043$ \\
         \cline{2-5}
        & $\frac{1}{6} \le \overline{\omega }_{rh} \le \frac{2}{3}$ & $0.968 \le n_s \le 0.972$ & $52.43 \le N_* \le 63.08$ & $0.043  \ge r \ge 0.033$ \\
        \cline{2-5}
        & $\frac{2}{3} \le \overline{\omega }_{rh} \le 1$ & $0.972 \le n_s \le 0.974$ & $63.08 \le N_* \le 67.21$ & $0.033 \ge r \ge 0.030$ \\
        \hlineB{4}
         \multirow{4}{*}{u = $25 M_P$} 
       & $-\frac{1}{3} \le \overline{\omega }_{rh} \le 0$ & $0.940 \le n_s \le 0.966$ & $25.54 \le N_* \le 46.55$ & $0.123 \ge r \ge 0.061 $ \\
         \cline{2-5}
        & $0 \le \overline{\omega }_{rh} \le \frac{1}{6}$ & $0.966 \le n_s \le 0.969$ & $46.55 \le N_* \le 52.51$ & $0.061  \ge r \ge 0.053$ \\
         \cline{2-5}
        & $\frac{1}{6} \le \overline{\omega }_{rh} \le \frac{2}{3}$ & $0.969 \le n_s \le 0.975$ & $52.51 \le N_* \le 63.18$ & $0.053  \ge r \ge 0.042$ \\
        \cline{2-5}
        & $\frac{2}{3} \le \overline{\omega }_{rh} \le 1$ & $0.975 \le n_s \le 0.976$ & $63.18 \le N_* \le 67.32$ & $0.042 \ge r \ge 0.038$ \\
         \hline
    \end{tabular}
   
    \label{T1}
\end{table}
\begin{table}
\caption{The allowed ranges of $n_s$, \(N_*\) and $r$ for 3 different choices of ${\mu}$ for normal hilltop inflation by demanding $p=3$ and $T_{rh} \geq 100GeV$}
    \centering
    \begin{tabular}{|c|c|c|c|c|c|}
        \hline
          n = 3 & Effective equation of state & $n_s$ & $ N_*$ & $r$ \\
         \hlineB{4}
         \multirow{5}{*}{u = $8 M_P$}
        & $-\frac{1}{3} \le \overline{\omega }_{rh} \le 0$ & $0.914 \le n_s \le 0.945$ & $25.20 \le N_* \le 45.90$ & $0.027   \ge r \ge 0.007$  \\
         \cline{2-5}
        & $0 \le \overline{\omega }_{rh} \le \frac{1}{6}$ & $0.945 \le n_s \le 0.950$ & $45.90 \le N_* \le 51.74$ & $0.007 \ge r \ge 0.005 $ \\
         \cline{2-5}
        & $\frac{1}{6} \le \overline{\omega }_{rh} \le \frac{2}{3}$ & $0.950 \le n_s \le 0.956$ & $51.74 \le N_* \le 62.20$ & $0.005 \ge r \ge 0.003$ \\
        \cline{2-5}
        & $\frac{2}{3} \le \overline{\omega }_{rh} \le 1$ & $0.956 \le n_s \le 0.958$ & $62.20 \le N_* \le 66.25$ & $0.003 \ge r \ge 0.025$ \\
        \hlineB{4}
        \multirow{4}{*}{u = $15 M_P$}
        & $-\frac{1}{3} \le \overline{\omega }_{rh} \le 0$ & $0.933 \le n_s \le 0.960$ & $25.38 \le N_* \le 46.27$ & $0.069 \ge r \ge 0.027 $ \\
         \cline{2-5}
        & $0 \le \overline{\omega }_{rh} \le \frac{1}{6}$ & $0.960 \le n_s \le 0.963$ & $46.27 \le N_* \le 52.18$ & $0.027  \ge r \ge 0.022$ \\
         \cline{2-5}
        & $\frac{1}{6} \le \overline{\omega }_{rh} \le \frac{2}{3}$ & $0.963 \le n_s \le 0.968$ & $52.18 \le N_* \le 62..78$ & $0.022  \ge r \ge 0.015$ \\
        \cline{2-5}
        & $\frac{2}{3} \le \overline{\omega }_{rh} \le 1$ & $0.968 \le n_s \le 0.970$ & $62.78 \le N_* \le 66.88$ & $0.015 \ge r \ge 0.014$ \\
        \hlineB{4}
         \multirow{4}{*}{u = $25 M_P$} 
         & $-\frac{1}{3} \le \overline{\omega }_{rh} \le 0$ & $0.939 \le n_s \le 0.965$ & $25.47 \le N_* \le 46.44$ & $0.099 \ge r \ge 0.045 $ \\
         \cline{2-5}
        & $0 \le \overline{\omega }_{rh} \le \frac{1}{6}$ & $0.965 \le n_s \le 0.969$ & $46.44 \le N_* \le 52.38$ & $0.045  \ge r \ge 0.038$ \\
         \cline{2-5}
        & $\frac{1}{6} \le \overline{\omega }_{rh} \le \frac{2}{3}$ & $0.969 \le n_s \le 0.973$ & $52.38 \le N_* \le 63.03$ & $0.038  \ge r \ge 0.029$ \\
        \cline{2-5}
        & $\frac{2}{3} \le \overline{\omega }_{rh} \le 1$ & $0.973 \le n_s \le 0.975$ & $63.03 \le N_* \le 67.16$ & $0.029 \ge r \ge 0.027$ \\
         \hline
    \end{tabular}
   
    \label{T2}
\end{table}

\begin{table}
\caption{The allowed ranges of $n_s$, \(N_*\) and $r$ for 3 different choices of ${\mu}$ for normal hilltop inflation by demanding $p=4$ and $T_{rh} \geq 100GeV$}
    \centering
    \begin{tabular}{|c|c|c|c|c|c|}
        \hline
          n = 4 & Effective equation of state & $n_s$ & $ N_*$ & $r$ \\
         \hlineB{4}
         \multirow{5}{*}{u = $7 M_P$}
        & $-\frac{1}{3} \le \overline{\omega }_{rh} \le 0$ & $0.904 \le n_s \le 0.946$ & $25.05 \le N_* \le 45.67$ & $0.013   \ge r \ge 0.003$  \\
         \cline{2-5}
        & $0 \le \overline{\omega }_{rh} \le \frac{1}{6}$ & $0.946 \le n_s \le 0.950$ & $45.67 \le N_* \le 51.49$ & $0.003 \ge r \ge 0.002 $ \\
         \cline{2-5}
        & $\frac{1}{6} \le \overline{\omega }_{rh} \le \frac{2}{3}$ & $0.950 \le n_s \le 0.958$ & $51.49 \le N_* \le 61.93$ & $0.002 \ge r \ge 0.0015$ \\
        \cline{2-5}
        & $\frac{2}{3} \le \overline{\omega }_{rh} \le 1$ & $0.958 \le n_s \le 0.961$ & $61.93 \le N_* \le 65.96$ & $0.0015 \ge r \ge 0.0013$ \\
        \hlineB{4}
        \multirow{4}{*}{u = $15 M_P$}
        & $-\frac{1}{3} \le \overline{\omega }_{rh} \le 0$ & $0.930 \le n_s \le 0.958$ & $25.29 \le N_* \le 46.14$ & $0.051 \ge r \ge 0.018 $ \\
         \cline{2-5}
        & $0 \le \overline{\omega }_{rh} \le \frac{1}{6}$ & $0.958 \le n_s \le 0.962$ & $46.14 \le N_* \le 52.04$ & $0.018  \ge r \ge 0.015$ \\
         \cline{2-5}
        & $\frac{1}{6} \le \overline{\omega }_{rh} \le \frac{2}{3}$ & $0.962 \le n_s \le 0.968$ & $52.04 \le N_* \le 62.62$ & $0.015  \ge r \ge 0.010$ \\
        \cline{2-5}
        & $\frac{2}{3} \le \overline{\omega }_{rh} \le 1$ & $0.968 \le n_s \le 0.969$ & $62.62 \le N_* \le 66.72$ & $0.010 \ge r \ge 0.009$ \\
        \hlineB{4}
         \multirow{4}{*}{u = $30 M_P$}
        & $-\frac{1}{3} \le \overline{\omega }_{rh} \le 0$ & $0.938 \le n_s \le 0.965$ & $25.44 \le N_* \le 46.40$ & $0.092 \ge r \ge 0.041 $ \\
         \cline{2-5}
        & $0 \le \overline{\omega }_{rh} \le \frac{1}{6}$ & $0.965 \le n_s \le 0.969$ & $46.40 \le N_* \le 52.34$ & $0.041  \ge r \ge 0.035$ \\
         \cline{2-5}
        & $\frac{1}{6} \le \overline{\omega }_{rh} \le \frac{2}{3}$ & $0.969 \le n_s \le 0.973$ & $52.34 \le N_* \le 62.99$ & $0.035  \ge r \ge 0.027$ \\
        \cline{2-5}
        & $\frac{2}{3} \le \overline{\omega }_{rh} \le 1$ & $0.973 \le n_s \le 0.975$ & $62.99 \le N_* \le 67.12$ & $0.027 \ge r \ge 0.024$ \\
         \hline
    \end{tabular}
   
    \label{T3}
\end{table}
 \begin{figure}
      \centering
 \includegraphics[width=\textwidth]{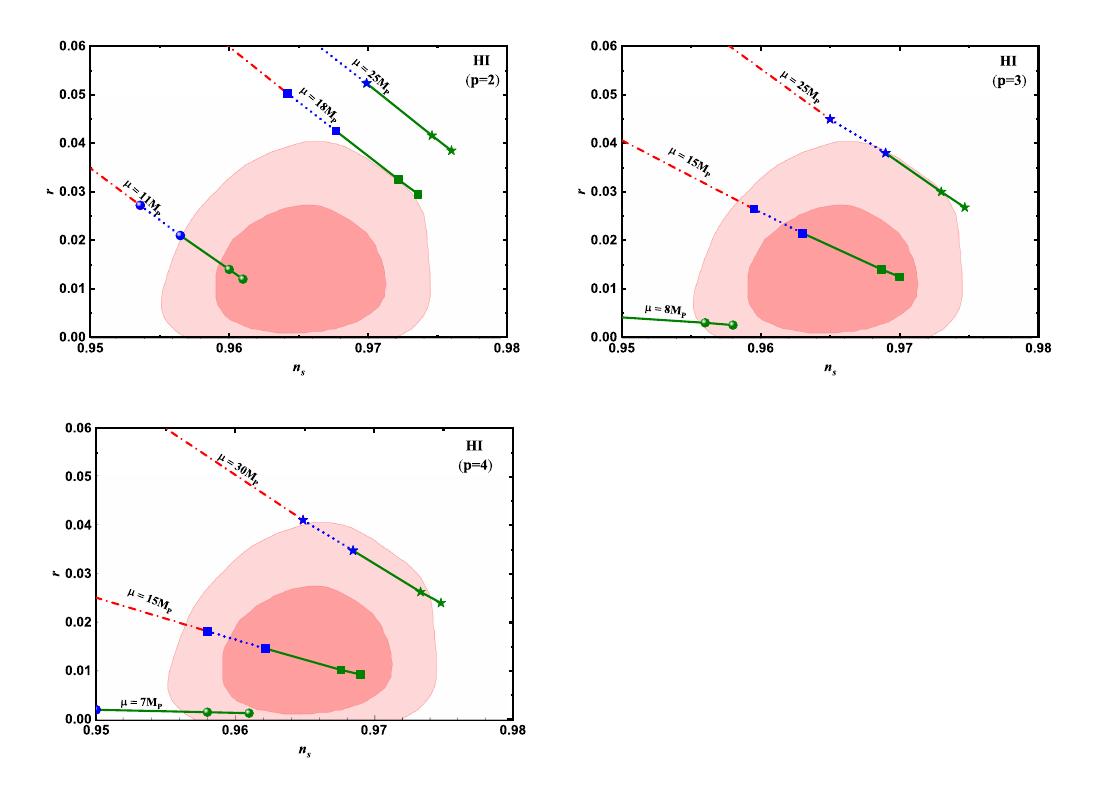}
      \caption{The {r} versus \(n_s\) plot for 3 different values of power index(p) in case of hilltop inflation with different {$\mu$} and \(\overline{\omega }_{\text{rh} }\): $-\frac{1}{3} \le \overline{\omega }_{rh} < 0$(dotted-dash red), $0 \le \overline{\omega }_{rh} \le \frac{1}{6}$(blue dotted), $\frac{1}{6} < \overline{\omega }_{rh} \le 1$(solid green). The light and dark peach shaded background contours are from Planck
TT,TE,EE+lowE+lensing+BK18+BAO observations \cite{tristram2022improved}.}
      \label{F6}
  \end{figure}
    \end{subsection}
   \begin{subsection}{Mutated Hilltop Inflation(MHI)}\label{S3.2}
The mutated hilltop potential is a variation on the hilltop inflation model, and its distinguishing feature is that the flat potential is modified by a hyperbolic function, whose power series expansion contains an infinite number of terms, rather than a simple addition of few terms, therefore making the model more accurate and concrete at same time. The mutated hilltop potential is given as \cite{pal_mutated_2010,martin_encyclopaedia_2013,,pal_mutated_2018,pal2012semi}
\begin{equation}\label{26}
V = M^4[1- {\text{sech}}(\frac{\phi}{\mu})],
\end{equation}
where M is the normalization term and $\mu$ is the mass scale. Using eqs. (\ref{4}) and (\ref{5}), the hubble parameter and slow-roll parameters for MHI potential can be expressed as\\
\begin{equation}\label{27}
H^2 = \frac{M^4[1- {\text{sech}}(\frac{\phi}{\mu})]]}{3M_P^2}.
\end{equation} 
\begin{equation}\label{28}
\epsilon =\frac{M_P^2}{2 u^2}\coth \left(\frac{\phi }{2 u}\right)^2 \text{sech}\left(\frac{\phi }{\mu}\right)^2.
\end{equation}
\begin{equation}\label{29}
\eta =-\frac{M_P^2}{4 u^2}\left(-3+\cosh \left(\frac{2 \phi }{\mu}\right)\right) \text{csch}\left(\frac{\phi }{2 u}\right)^2 \text{sech}\left(\frac{\phi
}{\mu}\right)^2.
\end{equation}
 The number of e-folds left after the pivot scale \(k_*\) crosses the Hubble radius
\begin{equation}\label{30}
\text{N}_*\simeq \frac{1}{M_P^2}\int_{\phi _{\text{end}}}^{\phi _*} \frac{V}{V'} \, d\phi \\
\\ = \frac{u^2 \left(-\cosh \left(\frac{\phi _{\text{end}}}{\mu}\right)+\cosh \left(\frac{\phi _*}{\mu}\right)+2 \ln \left[\cosh \left(\frac{\phi _{\text{end}}}{2
u}\right) \text{sech}\left(\frac{\phi _*}{2 u}\right)\right]\right)}{M_P^2}.
\end{equation}
Using the eqs. (\ref{19}) and (\ref{20})\ for $\epsilon $ and $\eta $ in eq. (\ref{7}) and  eq. (\ref{8}), the scalar spectral index and tensor-to-scalar ratio \(r=16\epsilon\) can be given as
\begin{equation}\label{31}
n_s=1-\frac{M_P^2}{2 u^2}\text{csch}\left(\frac{\phi _*}{2 u}\right)^2 \left(1+3 \text{sech}\left(\frac{\phi _*}{\mu}\right)+\tanh \left(\frac{\phi
_*}{\mu}\right)^2\right).
\end{equation}
\begin{equation}\label{32}
r=\frac{8M_P^2}{u^2}\coth \left(\frac{\phi _*}{2 u}\right)^2 \text{sech}\left(\frac{\phi _*}{\mu}\right)^2.
\end{equation}
The variation of \(N_*\) with \(n_{s\text{  }}\) is shown in figure (\ref{F7}b).
Additionally, this model gives the relation
\begin{equation}\label{33}
H_*= 2\pi  M_P^2 \sqrt{\frac{\coth \left(\frac{\phi _*}{2 u}\right)^2 \text{sech}\left(\frac{\phi _*}{\mu}\right)^2 A_s }{u^2}}.
\end{equation}
This, along with eq. (\ref{27}) and the condition defining the inflationary end, gives the value \
\begin{equation}\label{34}
V_{\text{end}}= 3H_*^2 M_P^2 \frac{1-\text{sech} \left(\frac{\phi _{\text{end}}}{\mu}\right)}{1-\text{sech} \left(\frac{\phi _*}{\mu}\right)}.
\end{equation}
\begin{figure}[!h]
    \centering
\includegraphics[width=\textwidth]{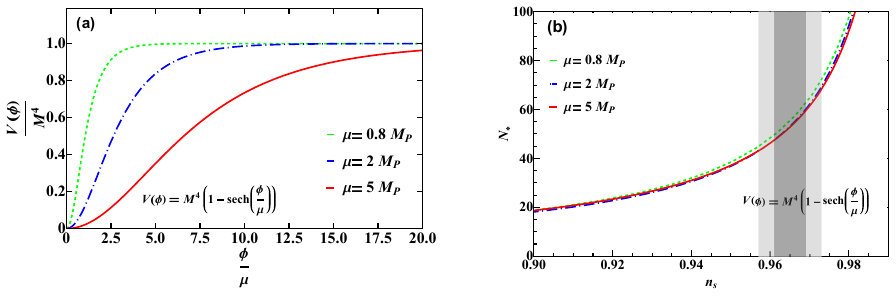}
    \caption{The variation of (a) Potential with $\frac{\phi}{M_P}$ and (b) \(N_*\) as a function of spectral index (\(n_{s }\)) for mutated hilltop inflation \(V = M^4[1- {\text{sech}}(\frac{\phi}{\mu})]\) for different $\mu$ values: $\mu= 0.8M_P$(medium dashed green), \(\mu= 2M_P\)(dot dashed blue), \(\mu= 5M_P\)(solid red). The shaded regions are same as in figure \ref{F2}}.
    \label{F7}
\end{figure}
Now, \(N_*\), $n_s$, \(H_{\text{*}}\) and \(V_{\text{end}}\) from eqs. (\ref{30}), (\ref{31}), (\ref{33}) and (\ref{34}) are all inserted in eqs. (\ref{15}) and (\ref{16}) to get plots of \(N_{\text{rh}}\) and \(T_{\text{rh}}\) as a function of \(n_s\) and are displayed in figure \ref{F8} for 3 distinct $\mu$ values taken. The \(T_{\text{rh}}\) vs \(n_s\) plots of figure \ref{F8} illustrates that for $\mu < 5M_P$ the curves corresponding to ($0 \le \overline{\omega }_{rh} \le 1$) lies well within the observable \(n_s\) bounds and as we move towards $\mu < M_P$ the $ \overline{\omega }_{rh} = 0$ curve starts slightly shifting outside the bounded region for lower \(T_{\text{rh}}\) values. Figure \ref{F8} also illustrates points where all curves for different $ \overline{\omega }_{rh}$ converges these are the points when instantaneous reheating occurs $(N \to 0)$, the temperature is maximum at these points and is independent of $ \overline{\omega }_{rh} $.\\
Moving further demanding \(T_{\text{rh}} \ge 100\) GeV, we have obtained allowed range of \(n_s\) and reflected it on eq. (\ref{30}) and (\ref{32}) to get bounds on \(N_*\) and r for 3 different $\mu$ values and are presented in table \ref{T4}. From table \ref{T4} it can be seen that for all $\mu$ values studied \(n_s\) lies well within the compatible range for $ \overline{\omega }_{rh} > 0$ and if we consider $\overline{\omega }_{rh} \le 1$, \(N_*\) is allowed to take values around 65 to 67.\\
The r versus $n_s$ predictions from table \ref{T4} for different $\mu$ over a range of  $\omega$ are presented in figure \ref{F9} with observational data contours in background. It can be seen from figure \ref{F9} that there is an upper bound on $\mu$ only below which MHI model is consistent with Planck's data. There is no $\mu_{min}$ observed for MHI, unlike the case of normal hilltop inflation. It was also observed that at lower values  of $\mu$ (especially $\mu < M_P$), we got relatively lower tensor to scalar ratio in comparison to normal hilltop inflation without any incompatibility of $n_s$ with observational data.
\begin{figure}
    \centering
\includegraphics[width=\textwidth]{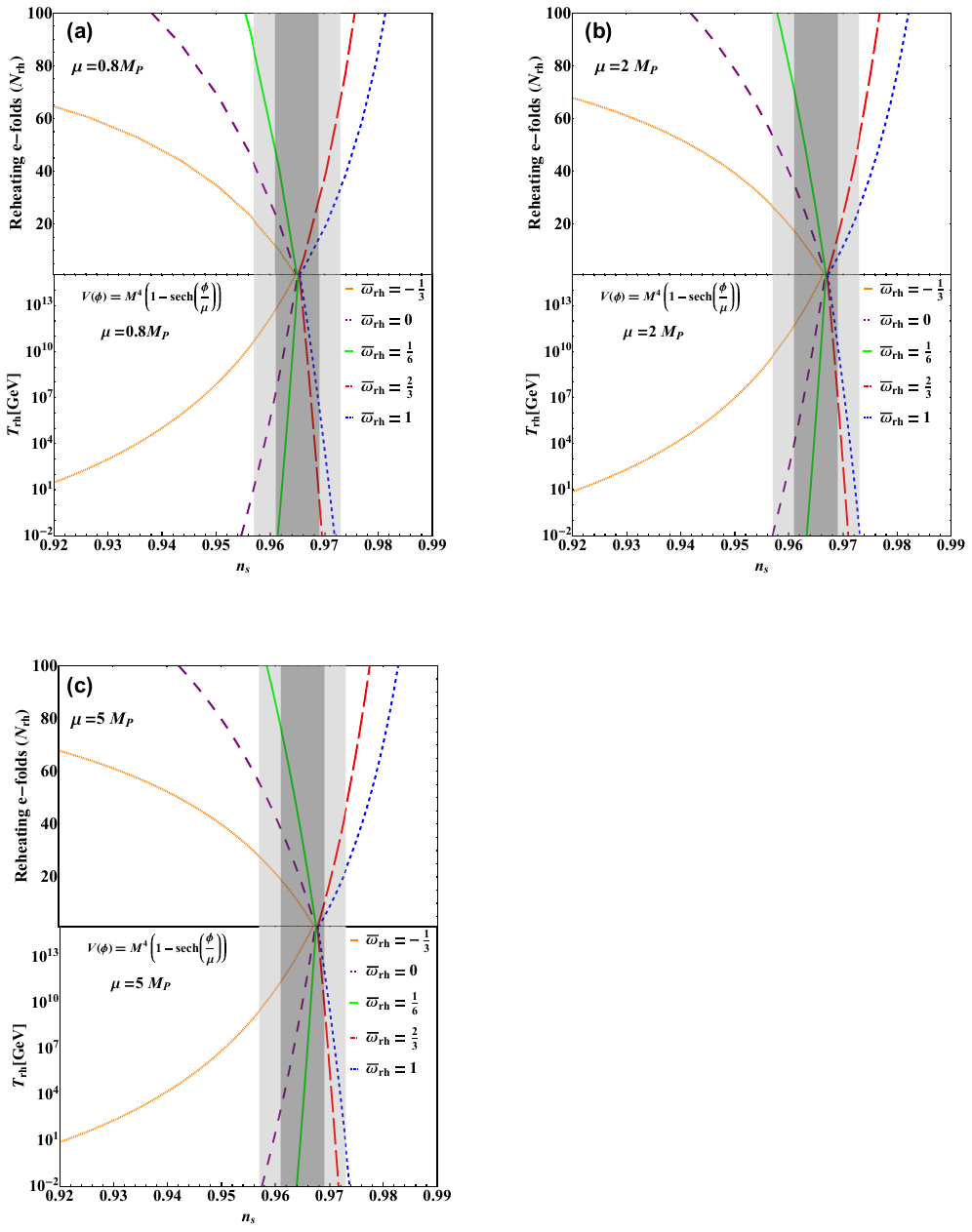}
    \caption{The variation of \(N_{\text{rh}}\) and \(T_{\text{rh}}\)  with spectral index (\(n_{s }\)) for mutated hilltop inflation \(V = M^4[1- {\text{sech}}(\frac{\phi}{\mu})]\) for different values of ${\mu}$ and \(\overline{\omega }_{\text{rh}}\). The colour coding and grey shadings are same as in figure \ref{F3}}
    \label{F8}
\end{figure}
\begin{table}[!t]
\caption{The allowed ranges of $n_s$, \(N_*\) and $r$ for 3 different choices of ${\mu}$ for mutated hilltop inflation by demanding $T_{rh} \geq 100GeV$}
    \centering
    \begin{tabular}{|c|c|c|c|c|c|}
        \hline
         \multirow{5}{*}{u = $0.8 M_P$} & Effective equation of state & $n_s$ & $ N_*$ & $r$ \\
        \cline{2-5}
        & $-\frac{1}{3} \le \overline{\omega }_{rh} \le 0$ & $0.924 \le n_s \le 0.957$ & $24.90 \le N_* \le 45.50$ & $6.95 \times 10^{-3}  \ge r \ge 2.21 \times 10^{-3}$ \\
         \cline{2-5}
        & $0 \le \overline{\omega }_{rh} \le \frac{1}{6}$ & $0.957 \le n_s \le 0.962$ & $45.50 \le N_* \le 51.33$ & $2.21 \times 10^{-3}  \ge r \ge 1.75 \times 10^{-3}$ \\
         \cline{2-5}
        & $\frac{1}{6} \le \overline{\omega }_{rh} \le \frac{2}{3}$ & $0.962 \le n_s \le 0.969$ & $51.33 \le N_* \le 61.73$ & $1.75 \times 10^{-3}  \ge r \ge 1.22 \times 10^{-3}$ \\
        \cline{2-5}
        & $\frac{2}{3} \le \overline{\omega }_{rh} \le 1$ & $0.969 \le n_s \le 0.970$ & $61.73 \le N_* \le 65.35 $ & $1.22 \times 10^{-3} \ge r \ge 1.09 \times 10^{-3}$ \\
        \hlineB{4}
         \multirow{4}{*}{u = $2 M_P$} 
        & $-\frac{1}{3} \le \overline{\omega }_{rh} \le 0$ & $0.928 \le n_s \le 0.959$ & $25.32 \le N_* \le 46.06$ & $0.031  \ge r \ge 0.011$ \\
         \cline{2-5}
        & $0 \le \overline{\omega }_{rh} \le \frac{1}{6}$ & $0.959 \le n_s \le 0.964$ & $46.06 \le N_* \le 51.93$ & $0.011  \ge r \ge 0.009$ \\
         \cline{2-5}
        & $\frac{1}{6} \le \overline{\omega }_{rh} \le \frac{2}{3}$ & $0.964 \le n_s \le 0.970$ & $51.93 \le N_* \le 62.47$ & $0.009  \ge r \ge 0.006$ \\
        \cline{2-5}
        & $\frac{2}{3} \le \overline{\omega }_{rh} \le 1$ & $0.970 \le n_s \le 0.972$ & $62.47 \le N_* \le 66.55$ & $0.006  \ge r \ge 0.005$ \\
         \hlineB{4}
         \multirow{4}{*}{u = $5 M_P$} 
        & $-\frac{1}{3} \le \overline{\omega }_{rh} \le 0$ & $0.928 \le n_s \le 0.960$ & $25.89 \le N_* \le 46.69$ & $0.010  \ge r \ge 0.039$ \\
         \cline{2-5}
        & $0 \le \overline{\omega }_{rh} \le \frac{1}{6}$ & $0.960 \le n_s \le 0.965$ & $46.69 \le N_* \le 52.58$ & $0.039  \ge r \ge 0.032$ \\
         \cline{2-5}
        & $\frac{1}{6} \le \overline{\omega }_{rh} \le \frac{2}{3}$ & $0.965 \le n_s \le 0.971$ & $52.58 \le N_* \le 63.13$ & $0.032  \ge r \ge 0.024$ \\
        \cline{2-5}
        & $\frac{2}{3} \le \overline{\omega }_{rh} \le 1$ & $0.971 \le n_s \le 0.973$ & $63.13 \le N_* \le 67.23$ & $0.024  \ge r \ge 0.022$ \\
        \hline
    \end{tabular}
   
    \label{T4}
    \end{table}
\begin{figure}
 \centering
    \includegraphics[width=0.6\textwidth]{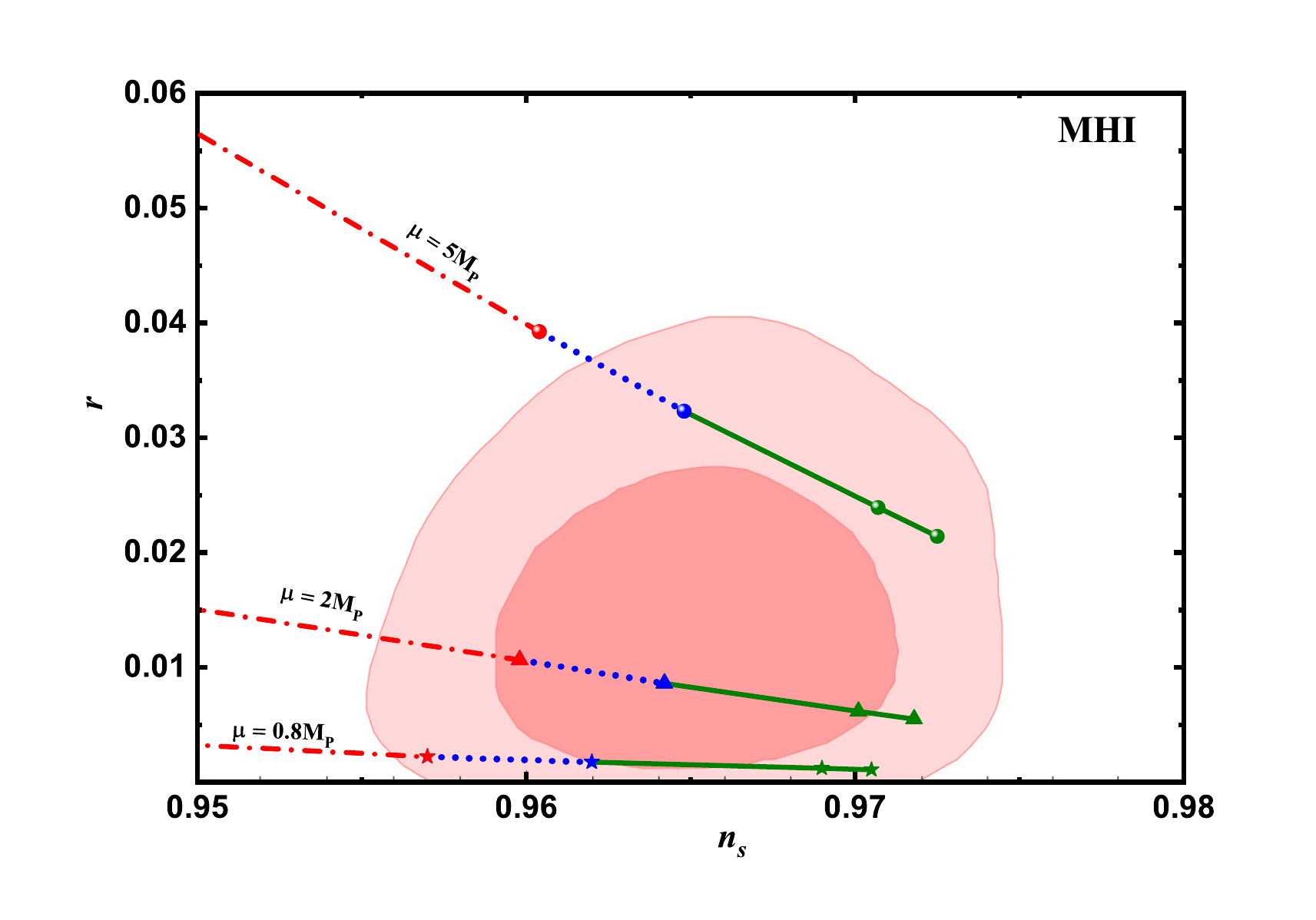}
    \caption{The {r} versus \(n_s\) plot for mutated hilltop inflation for 3 different {$\mu$} values over a range of \(\overline{\omega }_{\text{rh} }\). The colour codes and background contours are identical to figure \ref{F6}.} 
\label{F9}
\end{figure}
\end{subsection} 
\end{section}
\begin{subsection}{Discussion and Conclusion}\label{S4}
In this study, we have done the reheating analysis of hilltop and mutated hilltop models of inflation with model parameter $\mu$ in light of Planck 2018 + BK18 observations. We have carried out our study by examining how the reheating parameters, duration and temperature of reheating, \(N_{\text{rh}}\) and \(T_{\text{rh}}\) vary with scalar spectral index over a range of \(\overline{\omega }_{\text{rh}}\). By demanding \(T_{\text{rh}}> 100\) GeV and allowing \(\overline{\omega }_{\text{rh}}\) to vary in the range ($-\frac{1}{3} \le \overline{\omega }_{\text{rh}} \le 1$), we tried to constrain the model parameter space.\\
We first reexamined the normal hilltop inflation using the most recent observational data. Using the reheating conditions, we observed that for normal hilltop inflation, there is a range of $\mu$ values for each power index for which model is consistent with data and with increasing value of power index(p) this $\mu$ range becomes wider. It can also be seen that below a minimum $\mu$ there is inconsistency with observational $n_s$ value and as the power index(p) increase this minimum $\mu$ shifts towards lower $\mu$ values. The tensor to scalar ratio for this model lies in the range $r \sim 10^{-2} $. \\
Moving further we have done the reheating study of mutated hilltop inflation. On doing the reheating study of MHI we observed that if we consider the whole range of \(T_{\text{rh}}\) i.e. (\(10^{-2}GeV \le T_{\text{rh}} \le 10^{16}GeV\)) then $\mu = 2M_P$ shows consistency with Planck $2\sigma$ bound on $n_s$ for the widest range of $\overline{\omega }_{rh}$ i.e. ($0 \le \overline{\omega }_{rh} \le 1$).
On imposing the conditions \(T_{\text{rh}}> 100\) GeV, we found that there is an upper bound on $\mu$ only below which MHI model is consistent with data. The $\mu$ values around $2M_P$ have the widest $\overline{\omega }_{rh}$ range for which they are compatible with data making them the most favourable parameter space choice. There is no $\mu_{min}$ observed for MHI, unlike the case of normal hilltop inflation. It was also observed that at lower values  of $\mu$ (especially $\mu < M_P$), we got relatively lower tensor to scalar ratio in comparison to normal hilltop inflation without any incompatibility of $n_s$ with observational data making it a
better choice in accordance to recent and future studies.
\end{subsection}
\printbibliography
\end{document}